\documentclass[preprint,showpacs,preprintnumbers,amsmath,amssymb,superscriptaddress]{revtex4}

\usepackage{graphicx}
\usepackage{dcolumn}
\usepackage{bm}

\begin{document}


\title{Hot electron cooling by acoustic phonons in graphene}

\author{A.C. Betz}
\affiliation{Laboratoire Pierre Aigrain, ENS-CNRS UMR 8551,
Universit\'es P. et M. Curie and Paris-Diderot,
24, rue Lhomond, 75231 Paris Cedex 05, France}
\author{F. Vialla}
\affiliation{Laboratoire Pierre Aigrain, ENS-CNRS UMR 8551,
Universit\'es P. et M. Curie and Paris-Diderot,
24, rue Lhomond, 75231 Paris Cedex 05, France}
\author{D. Brunel}
\affiliation{Laboratoire Pierre Aigrain, ENS-CNRS UMR 8551,
Universit\'es P. et M. Curie and Paris-Diderot,
24, rue Lhomond, 75231 Paris Cedex 05, France}
\author{C. Voisin}
\affiliation{Laboratoire Pierre Aigrain, ENS-CNRS UMR 8551,
Universit\'es P. et M. Curie and Paris-Diderot,
24, rue Lhomond, 75231 Paris Cedex 05, France}
\author{M. Picher}
\affiliation{Laboratoire de Photonique et Nanostructures,
CNRS-UPR20 CNRS, Route de Nozay, 91460 Marcoussis
Cedex, France.}
\author{A. Cavanna}
\affiliation{Laboratoire de Photonique et Nanostructures,
CNRS-UPR20 CNRS, Route de Nozay, 91460 Marcoussis
Cedex, France.}
\author{A. Madouri}
\affiliation{Laboratoire de Photonique et Nanostructures,
CNRS-UPR20 CNRS, Route de Nozay, 91460 Marcoussis
Cedex, France.}
\author{G. F\`eve}
\affiliation{Laboratoire Pierre Aigrain, ENS-CNRS UMR 8551,
Universit\'es P. et M. Curie and Paris-Diderot,
24, rue Lhomond, 75231 Paris Cedex 05, France}
\author{J.-M. Berroir}
\affiliation{Laboratoire Pierre Aigrain, ENS-CNRS UMR 8551,
Universit\'es P. et M. Curie and Paris-Diderot,
24, rue Lhomond, 75231 Paris Cedex 05, France}
\author{B. Pla\c{c}ais}
\email{bernard.placais@lpa.ens.fr}
\affiliation{Laboratoire Pierre Aigrain, ENS-CNRS UMR 8551,
Universit\'es P. et M. Curie and Paris-Diderot,
24, rue Lhomond, 75231 Paris Cedex 05, France}
\author{E. Pallecchi}
\affiliation{Laboratoire Pierre Aigrain, ENS-CNRS UMR 8551,
Universit\'es P. et M. Curie and Paris-Diderot,
24, rue Lhomond, 75231 Paris Cedex 05, France}

\date{\today}

\begin{abstract}
We have investigated the energy loss of hot electrons in metallic graphene by means of GHz noise thermometry at liquid helium temperature. We observe the electronic temperature $T\propto V$ at low bias in agreement with the heat diffusion to the leads described by the Wiedemann-Franz law. We report on  $T\propto\sqrt{V}$ behavior at high bias, which corresponds to a  $T^4$  dependence of the cooling power. This is the signature of a 2D acoustic phonon cooling mechanism. From a heat equation analysis of the two regimes we extract accurate values  of the electron-acoustic phonon coupling constant $\Sigma$ in monolayer graphene. Our measurements point to an important effect of lattice disorder in the reduction of $\Sigma$, not yet considered by theory. Moreover, our study provides a strong and firm support to the rising field of graphene bolometric detectors.

\end{abstract}

\maketitle

Many remarkable properties of graphene stem from the chiral Dirac-fermion nature of carriers and the associated semi-metallic density of states \cite{CastroRMP2009,DasSarmaRMP2010}. Graphene is also peculiar from the phonon point of view with a large optical phonon (OP) energy of $160$--$200\;\mathrm{meV}$ and a weak coupling of electrons to acoustic phonons (APs)  \cite{Piscanec2006PRL,Kubakaddi2009PRB,Bistritzer2009PRL,Viljas2010PRB}. The later is due to the non-ionic character of the lattice and the 2D nature of the phonons. Both effects conspire to the weak electron-phonon scattering responsible for the high intrinsic carrier mobility of graphene up to room temperature \cite{Dean2010natnano}.
The OP-scattering reappears however in the high-field transport, where OP emission  by energetic electrons gives rise to  current saturation \cite{Bareiro2009PRL,Perebeinos2010PRB}. By contrast, extrinsic disorder effects usually obscure AP-scattering which remains thus more elusive and  still debated \cite{Bolotin2008PRL}.
The AP-coupling becomes however prominent in heat transport as it constitutes the bottleneck in the cooling process of hot electrons to the substrate.  Heat transport experiments in graphene offer thus the possibility to characterize electron-AP coupling in detail.  On the application side, the weak AP coupling opens an avenue for the realization of fast and sensitive bolometric detectors working in a broad frequency spectrum ranging from microwave to optics \cite{Freitag2009nanoletters,Gabor2011Science,Vora2011arXiv,Yan2011arXiv,Fong2012arXiv}. Also in this respect an in-depth and direct characterization of hot-electron effects will prove useful.

In this letter we report on the interaction between electrons and acoustic phonons in mono-layer graphene. To this end, we have investigated the bias dependence of the electron temperature deduced from radio frequency (RF) shot noise measurements. Our experiment provides a direct access to the electron-phonon coupling constant.
We rely on the relation $S_I=4k_BT_e/R$ between the current noise spectrum $S_I$, the average electronic temperature $T_e$, and the sample resistance $R$. Johnson-noise thermometry has proven useful to study carbon nanotubes \cite{Wu2010APL,Chaste2010APL} and more recently bilayer-graphene \cite{Fay2011PRB}.
From an experimental point of view, we measure noise in the MHz to GHz band to overcome the environmental $1/f$ contribution while keeping quasi-stationary response conditions \cite{Pallecchi2011PRB}.
At sufficiently high bias, we find $T_e\propto \sqrt{V}$, where V is the applied voltage. This is a signature of 2D acoustic phonons. We observe it in diffusive samples where the $I-V$ characteristics are linear and optical phonon scattering is absent.
At low bias we recover the  $T_e\propto V$ behavior, expected for heat conduction to the contacts, and described by the Wiedemann-Franz law.
Both regimes are well explained by the heat equation. By fitting it to our measurements we extract the coupling constant $\Sigma$  as function of carrier density for samples on different substrates.

\begin{figure}[ttt]
\centerline{\includegraphics[scale=0.5]{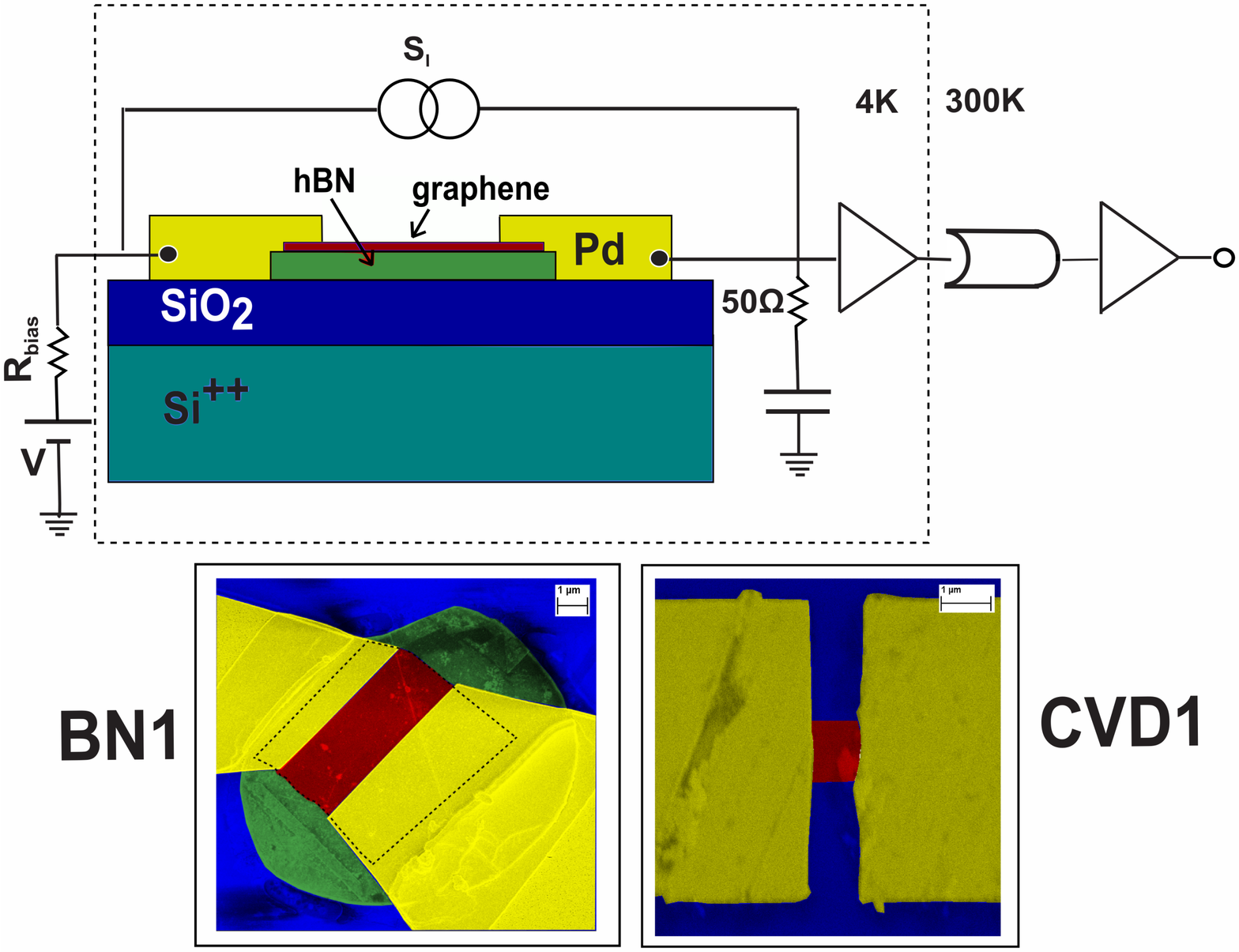}}
\caption{Shot noise measurement scheme (color online) with a sketch of the sample  the electrical environment. The samples and the cryogenic low-noise amplifier are cooled at $4.2\mathrm{K}$. The SEM pictures refer to samples BN1 and CVD1: BN1 consists of an exfoliated graphene flake (red), deposited on an exfoliated h-BN platelet (green), itself deposited on a Si/SiO$_2$ used as a back-gate. CVD1 was fabricated from CVD grown graphene deposited on Si/SiO$_2$. Both samples are contacted with Pd electrodes (yellow).}\label{chipoverview}
\end{figure}

\begin{table}[htbp]

\begin{tabular}{|l|c|c|} \tableline
\emph{ sample} & $L\times W$ ($\mathrm{\mu m^{2}}$) & R
($\mathrm{k\Omega}$ )  \\
\tableline
BN1 & $2.2\times5.7$ & $2.8$--$3.8$  \\ \tableline
BN2 & $2.2 \times 2.7$ & $1.3$--$2.3$     \\ \tableline
CVD1 & $1\times1$ & $ 1.67$    \\ \tableline
\end{tabular}
\caption{Characteristics of the graphene samples. $L$ is the sample length, $W$ the sample width, and $R$ the drain-source resistance.}

\label{table1}\end{table}
We  present data of three representative devices: two large exfoliated graphene flakes deposited on hexagonal boron nitride (hBN) and a smaller sample  on SiO$_2$ made from chemical vapor deposition (CVD)  graphene. The samples, labelled BN1, BN2, and  CVD1 respectively are sketched in Fig.\ref{chipoverview}.
Sample BN1 and BN2 consist of a stack of a monolayer graphene and a $30\;\mathrm{nm}$-thick hBN platelet on a Si/SiO$_2$ substrate.
The hBN platelet of approximately $10\;\mathrm{\mu m}$ diameter was produced by exfoliation of a high quality hBN powder (St. Gobain "Tr\`es BN"). The graphene flake was also obtained by exfoliation and placed on top of the hBN using a wet transfer technique \cite{TransferDelft}.
CVD1 was made from a graphene sheet grown on copper, transferred to the Si/SiO$_2$ \cite{Lia2009Science}.
The samples were produced by means of e-beam lithography and dry etching. Doped silicon was used as a back gate, with a $1\;\mathrm{\mu m}$ thick oxide,  and each sample was embedded in a coplanar wave-guide adapted for GHz frequencies \cite{Pallecchi2011APL}. Palladium metallizations ensure low ohmic contacts to graphene. Table \ref{table1} lists the samples' dimensions and electrical characteristics.

Measurements were carried out in liquid helium at $4.2\;\mathrm{K}$ to ensure a cold phonon bath.
The amplification line, of bandwidth  $0.8\;\mathrm{GHz}$, 
 is similar to the one in Ref.\,\cite{Chaste2010APL}. It includes a cryogenic low noise amplifier, calibrated against the poissonian noise of a $460\;\mathrm{Ohms}$ Al/AlOx/Al tunnel junction.
The samples' $I$-$V$ characteristics were measured via the voltage drop across the bias resistance.
In situ current annealing was performed  to improve the samples' electrical characteristics.
The carrier density $n_s$ of samples BN1 and BN2 are deduced from the gate dependence of the DC conductivity in a range of $V_g=\pm18\;\mathrm{V}$(BN1) and $V_g=\pm28\;\mathrm{V}$(BN2). Sample BN1 stayed p-doped over the whole gate range whereas BN2 could be swept through the neutrality point.
BN2's mobility is on the order of  $3000\;\mathrm{cm^{2}V^{-1}s^{-1}}$, about a factor 10 larger than that of BN1. The conductance of CVD1 was independent of $V_g$ which indicates a highly doped regime; we estimate the carrier concentration on the order of a few $10^{13}\;\mathrm{cm^{-2}}$ from Hall-bar measurements on similar CVD sheets.

\begin{figure}[ttt]
\centerline{\includegraphics[width=0.8\textwidth]{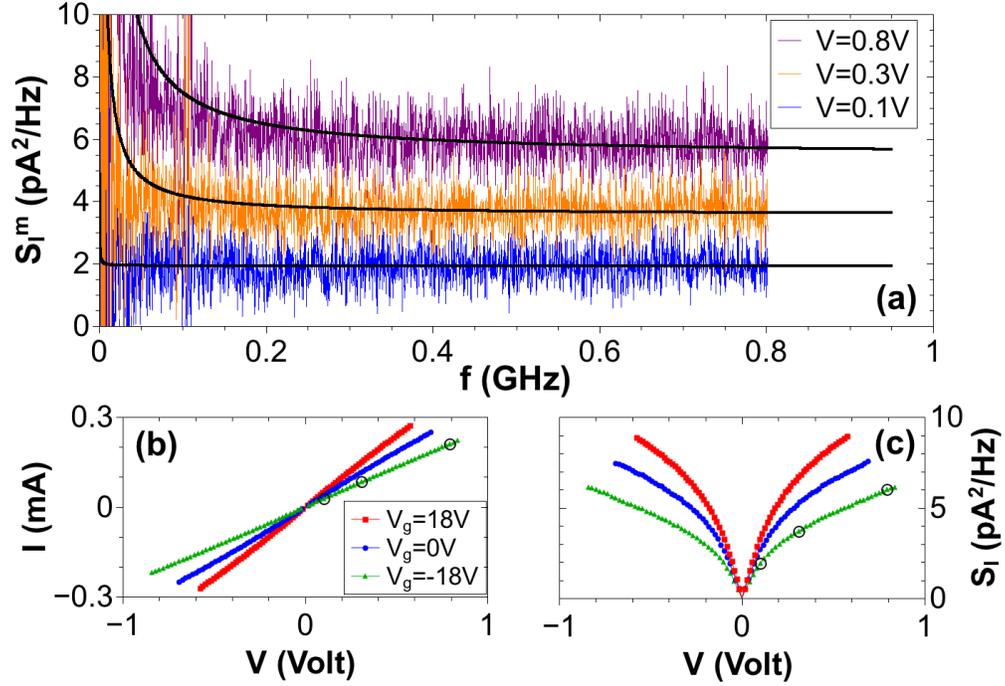}}
\caption{(a) Typical excess noise $S^{m}_I(V)$ spectra in sample BN1; it is a white noise with a superimposed $1/f$ contribution and fitted by  $S^{m}_I=S_{I}+C/f$ laws (solid lines). (b) and (c) show the $I(V)$ and $S_I(V)$ data for different gate voltages in sample BN1 from which we deduce $T_e$. The circles point out the $I(V)$ and $S_I(V)$ values of the spectra shown in (a).}\label{spectrum}
\end{figure}

\begin{figure}[ttt]
\centerline{\includegraphics[width=0.8\textwidth]{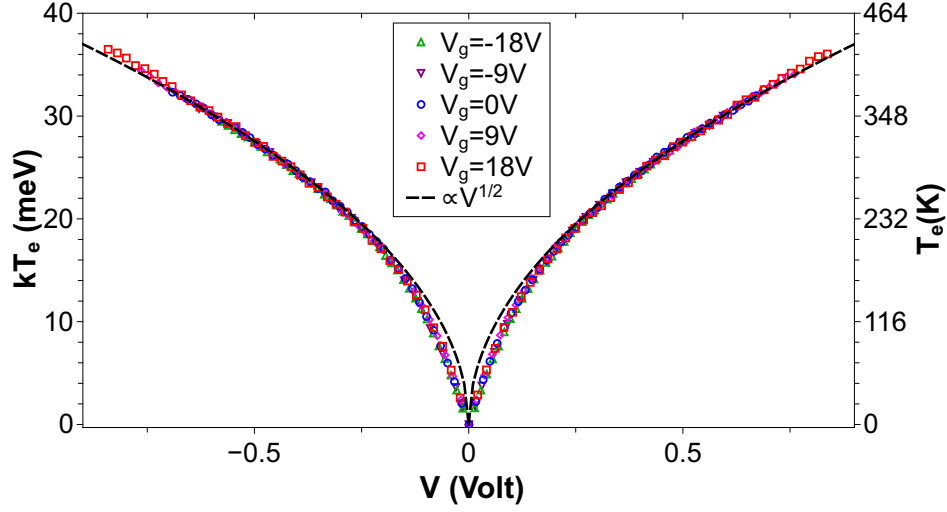}}
\caption{Electronic temperature in sample BN1 as function of voltage bias for a set of gate voltages. $T_e=S_I V/(4k_BI)$ is deduced from the $S_I(V)$ and $I(V)$ data shown in Fig.\ref{spectrum}(c) and (b). Unlike $S_I(V)$, $T_e(V)$ is nearly independent of  gate voltage and closely follows the $T_e\propto\sqrt{V}$ law (black dashed line) expected for 2D phonons. Deviations are observed at low bias where a $T_e\propto V$ behavior is found.}
\label{Fig3TeBN}\end{figure}

We turn now to the presentation of measurements. Fig.\ref{spectrum}(a) shows typical  current noise spectra for different bias voltages.  At low bias the spectrum is flat, whereas at higher bias a $1/f$ contribution clearly arises. Our broad-band setup allows us to quantitatively separate the white shot noise contribution $S_I$ from $1/f$ noise by fitting the spectra  with a $S_I+ C/f$ law. We find  $C\propto I^2$ in accordance with the Hoodge law \cite{Hoodge1981RPP}. This procedure is especially important for small samples where the $1/f$ contribution is larger.
Figs.\ref{spectrum}(b) and (c) show the current $I$ and the shot noise $S_I$ of sample BN1 as a function of bias  for three gate voltages.
We restrict our study to  biases  where $I$-$V$  characteristics are linear as the lack of current saturation is an indication of the absence of optical phonon scattering \cite{Bareiro2009PRL}. This is consistent with a transport scattering length ($\lesssim10\;\mathrm{nm}$) much smaller than the optical phonon scattering length ($\gtrsim200\;\mathrm{nm}$) \cite{Bareiro2009PRL}.
The shot noise is strongly sub-linear with respect to bias voltage and depends on gate voltage, with larger noise at higher carrier concentration.
We extract the bias dependence of the average electronic temperature $T_e(V)=S_IR/4k_B$ from the measured shot noise. The results for sample BN1 at five gate voltages are displayed in Fig.\ref{Fig3TeBN}. The electronic temperature increases with bias voltage and reaches several hundred Kelvin at high bias. This value is well above bath temperature but still below the optical phonon energy of $\simeq200\;\mathrm{meV}$.
We find that $T_e$ is weakly dependent on gate voltage and the different curves closely follow the $T_e\propto\sqrt{V}$ law (dashed line) predicted for 2D acoustic phonons.
Deviations are however observed at low bias where data rather follows a $T_e\propto V$ law. At the low end of the investigated bias range, the quasi-equilibrium approximation required to define an electron temperature may fail due to large electron-electron scattering lengths \cite{Steinbach1996PRL}. From a Fano factor analysis, we estimate this mesoscopic regime to arise below a few tens of $\mathrm{mV}$
\footnote{In our samples, the Fano factor extrapolates to a value close to 1/3 at vanishing voltage.}.

The electron temperature  is determined by the balance between the Joule heating and the cooling powers at play\,\cite{Nagaev1995PRB}. Joule heating per unit area is $P=V^2/RWL$, while cooling is  provided by two mechanisms: The first is electron heat diffusion to the leads, the second electron-AP interactions.
At sufficiently high bias the diffusive contribution can be neglected, i.e. the cooling power is given by $\Sigma (T_e^{\delta}-T_{ph}^{\delta})$, where $T_{ph}$ is the phonon temperature. In graphene $\delta=4$ has been proposed to account for the purely two dimensional APs \cite{Kubakaddi2009PRB,Viljas2010PRB}. In general  $\delta$ varies from $3$ to $6$ \cite{Giazotto2006RMP};  the lower value $\delta=3$ is observed in 1D systems such as carbon nanotubes \cite{Wu2010APL} and  $\delta=5$ is typically used for 3D metals \cite{Steinbach1996PRL,Huard2007PRB}.
For a quantitative analysis of our data, we solve the heat equation:
\begin{equation}
\frac{{\cal L}}{2R}\frac{L^2\partial^2T^2(x)}{\partial x^2}=-\frac{V^2}{R}+ LW\Sigma\left[T^4(x)-T^4_{ph}\right]\quad,
\label{BLequation}\end{equation}
where ${\cal L}=\pi^2k_B^2/3e^2$  is the Lorenz number and $x$ denotes the coordinate along the graphene channel.
We obtain an analytical solution in the case of cold contacts and cold phonon bath ($T_{ph}=0$). The solution depends on a single free parameter $\Sigma$, which sets the two characteristic scales of the system: the temperature $T_\Sigma=(V^2/RLW\Sigma)^{1/4}$ and the voltage $V_\Sigma={\cal L}/\sqrt{4RLW\Sigma}$.   $T_\Sigma=(P/\Sigma)^{1/4}$ is the maximum temperature  reached in the absence of electron heat conduction, whereas $V_\Sigma$ defines the crossover voltage between the electron cooling at low voltages and phonon cooling regime at high bias. The temperature profile is pseudo parabolic at low bias \cite{Nagaev1995PRB} and evolves toward a uniform temperature $T(x)=T_\Sigma$ at high bias. The spatial average $T_e=<T(x)>$ is then calculated numerically.

\begin{figure}[ttt]
\centerline{\includegraphics[width=0.7\textwidth]{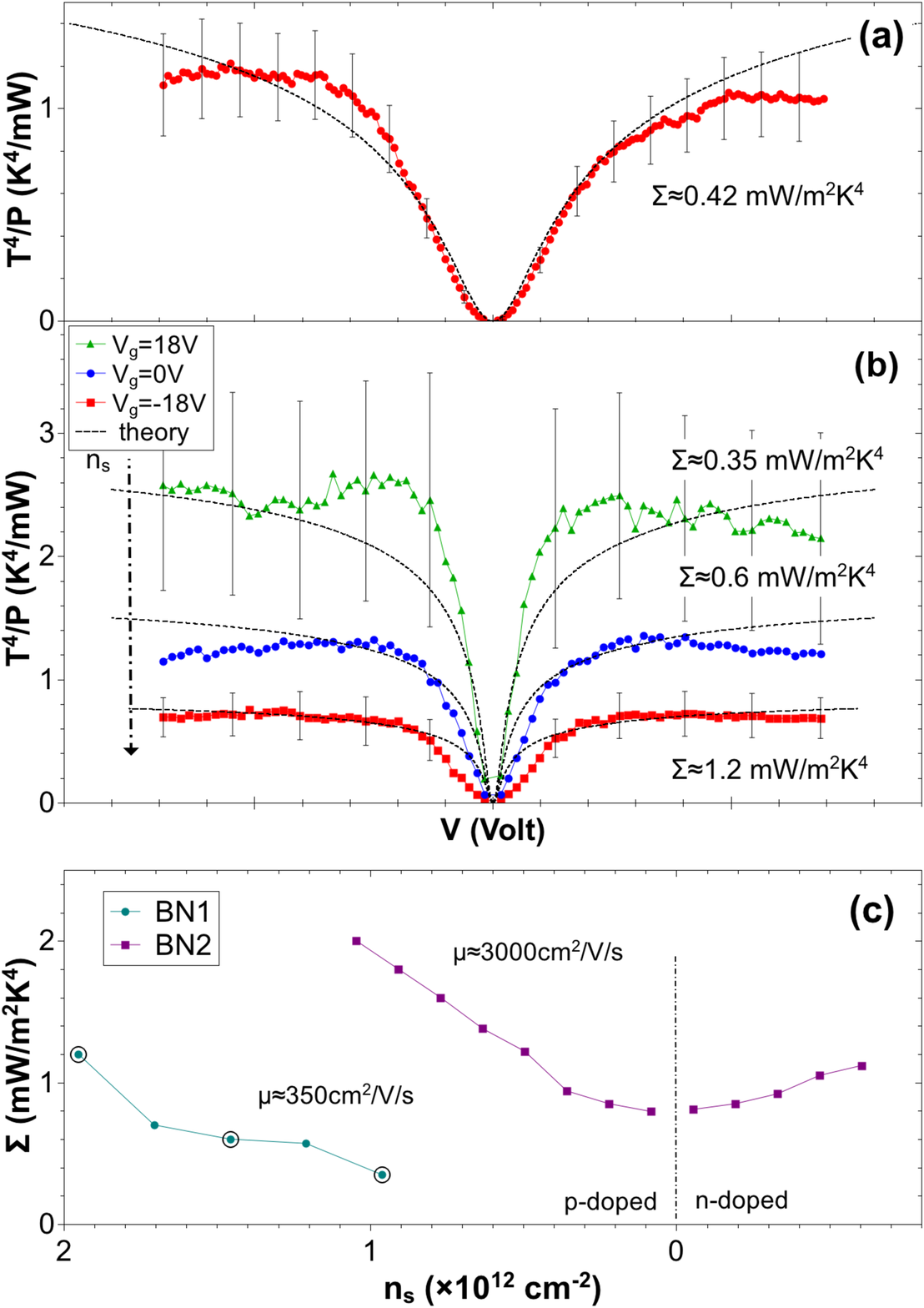}}
\caption{(a) and (b): Electron temperature of sample CVD1 (a) and BN1 (b) plotted as $T_e^4(V)/P$, where $P$ is the Joule heating per unit area $P=V^2/RLW$. The plateau at high bias is at a value $T_e^4/P\simeq1/\Sigma$. The dip at low V is due to electron heat diffusion to the leads.  Dashed lines are one-parameter fits with $\Sigma$ as free parameter. (c): $\Sigma$ as function of carrier density $n_s$ for samples BN1 and BN2.}\label{scalingBN}
\end{figure}

In Fig.\ref{scalingBN} we plot the temperature data in the form $T_e^4/P$ as a function of $V$ with the corresponding error bars. Due to a more pronounced $1/f$ contribution, the error increases with increasing bias. The $T_e^4/P$ representation allows for a better comparison with theory and puts the emphasis on the density dependence of  $\Sigma$. The  plateaus at high bias and the dips at low bias  reflect the $T_e\propto\sqrt{V}$  and $T_e\propto V$ regimes observed in the shot noise data (Fig.\ref{Fig3TeBN}). The measurements are well fitted by the solution of Eq.(\ref{BLequation}) taking $\Sigma$ as the only free parameter (solid lines). We find $\Sigma\lesssim 2\;\mathrm{mW/m^2/K^4}$  and a clear dependence on carrier density in BN samples, and   $\Sigma\simeq 0.42\;\mathrm{mW/m^2/K^4}$ in CVD1. Since the dip at low bias marks the effect of electron heat diffusion to the leads, the crossover is more pronounced in the shorter CVD1 sample.  This can be seen in  Fig.\ref{scalingBN}\,(a)  where the width of the dip is larger and agreement between data and theory is more quantitative. The fact that theory, with a single free parameter, accounts for both the plateau and the dip is a strong confirmation of the model and the underlying 2D acoustic phonon mechanism.
Fig.\ref{scalingBN}\,(c) shows  the carrier-density dependence  $\Sigma(n_s)$ and emphasizes its  sensitivity to carrier mobility and disorder. This observation is consistent with the smaller values of $\Sigma$ in the CVD1 sample.

\begin{figure}[ttt]
\centerline{\includegraphics[width=0.7\textwidth]{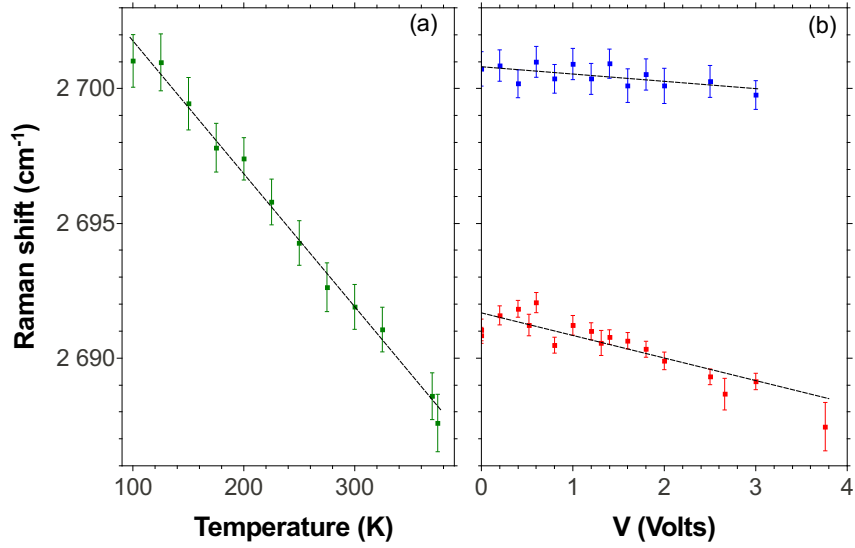}}
\caption{(a): Raman shift of the 2D-band of a graphene sample similar to CVD1,  as function of the cryostat temperature $T_0$.
The slope is $(-0.051\pm0.008)\;\mathrm{cm^{-1}K^{-1}}$. Excitation laser at $532\;\mathrm{nm}$ , $P=25\;\mathrm{kWcm^{-2}}$. No laser power dependence was observed in this range.
(b): Raman shift of the 2D band as function of the bias voltage for  $T_0=100\;\mathrm{K}$ (blue symbols) and $T_0=300\;\mathrm{K}$ (red symbols). The slopes are $(-0.3\pm0.3)\;\mathrm{cm^{-1}V^{-1}}$ and $(-0.9\pm0.3)\;\mathrm{cm^{-1}V^{-1}}$ respectively. The bias-induced phonon heating is therefore below $30\;\mathrm{K/V}$. Similar results and conclusions were drawn from measurements on the G band (not shown).
}
\label{Raman}\end{figure}

The afore mentioned $\Sigma T_e^4$ law for electron-AP cooling is similar to a $\Sigma_K T_{ph}^4$ law for phonon black body radiation to the substrate or the bath ($T_0$). This mechanism would occur in a hot phonon regime where $T_{ph}\simeq T_e\gg T_0$. However the corresponding coupling constant $\Sigma_K$ is about $3$--$4$ orders of magnitude larger than $\Sigma$ measured in our experiments \cite{Balandin2011nmat,Pobell2007}. Both mechanisms appear in series in the electron cooling process, thus giving rise to an effective coupling $\widetilde{\Sigma} =\Sigma \Sigma_K/(\Sigma+\Sigma_K)\simeq \Sigma$. This justifies our earlier hypothesis $T_{ph}^4\ll T_{e}^4$ in Eq.(\ref{BLequation}).
In order to confirm our cold acoustic phonon approximation we have performed Raman spectroscopy measurements in a similar sample.  As already reported by several groups \cite{CalizoAPL2007,Zhang2008JPCC}, we find that the position of the G (resp. 2D) band shows a downshift of $0.024 \;\mathrm{cm^{-1}/K}$ (resp. $0.051\;\mathrm{cm^{-1}/K}$) when increasing the cryostat temperature (Fig.\ref{Raman}(a)).
When the sample is biased, a similar downshift is observed, which is attributed to an increase of the AP bath temperature (Fig.\ref{Raman}(b)). This behavior was confirmed both at room temperature and at 100 K. We estimate that the corresponding increase of the AP bath temperature remains below $30\;\mathrm{K/V}$ in all cases.
As compared to electronic temperature elevations (Fig.\ref{Fig3TeBN}), these values  entail negligible corrections into Eq.(\ref{BLequation}).

We now come to the discussion of the results. According to theory\cite{Kubakaddi2009PRB,Viljas2010PRB}, longitudinal acoustic phonons (LA) are coupled to electrons via the deformation potential $D$ ($10$--$30\;\mathrm{eV}$). Theory predicts a cooling power  $P=\Sigma_{LA} T^4$   with  $\Sigma_{LA}\propto D^2\sqrt{n_s}$ in the non degenerate metallic regime and a different $P(T)$ law at high temperature. Taking $D=10\;\mathrm{eV}$, this gives $\Sigma_{LA}\simeq 10\sqrt{n_s/10^{12}}\;\mathrm{mW/m^2/K^4}$. Our experiment confirms the $P=\Sigma T^4$ dependence in the metallic regime, but we find a smaller coupling constant, $\Sigma=0.5\;\mathrm{mW/m^2/K^4}$ for BN1 and $\Sigma=2\;\mathrm{mW/m^2/K^4}$ for BN2 at $n_s=10^{12}\;\mathrm{cm^{-2}}$. The discrepancy in $\Sigma$ cannot be explained by experimental uncertainties. Our experiment suggest the effect of lattice disorder, which is the limiting factor for the mobility, as a possible cause of the reduction of the electron-phonon coupling. As a matter of fact, lattice disorder, which is known to affect phonon lifetime, is not taken into account by theory for the metallic regime \cite{Kubakaddi2009PRB,Viljas2010PRB}. In the high-temperature regime however, a recent theory  shows that disorder mediated phonon-phonon collisions may be important \cite{Song2011arXiv}.

In conclusion, we have studied the interaction between electrons and acoustic phonons in diffusive graphene by measuring the energy-loss of hot-electrons.
We find that the cooling power due to acoustic phonons follows a $\Sigma T^4$ law. We find that $\Sigma$ increases with carrier concentration, but the value, $\Sigma\lesssim 2 \;\mathrm{mW/m^2/K^4}$,  is approximately one order of magnitude smaller than predictions for LA-phonons. Our work is a motivation for further theoretical investigation that takes into account the role of lattice disorder on phonons.
Beside its implication for electron-phonon physics, our study of thermal noise is of direct relevance for the performance of graphene detectors.

\begin{acknowledgments}
We would like to thank R. Feirrera, B. Huard, T. Kontos, and F. Mauri for fruitful discussions, V. Bouchiat for additional CVD samples, S. Goossens, V. Calado, and L. Vandersypen for demonstration of their transfer technique, A. Denis for sample holder design, P. Morfin for engineering and S. Jang for reading the manuscript. St. Gobain kindly provided us with high quality hBN powder. The research has been supported by the contract ANR-2010-BLAN-MIGRAQUEL, SBPC and Cnano Gra-Fet-e.
\end{acknowledgments}

\end{document}